# Information Retrieval in Friction Stir Welding of Aluminum Alloys by using Natural Language Processing based Algorithms

Akshansh Mishra[1]

[1] Department of Chemistry, Materials and Chemical Engineering "Giulio Natta", Politecnico di Milano, Milan, Italy
[1]Orcid id: https://orcid.org/0000-0003-4939-359X

**Abstract:** Text summarization is a technique for condensing a big piece of text into a few key elements that give a general impression of the content. When someone requires a quick and precise summary of a large amount of information, it becomes vital. If done manually, summarizing text can be costly and time-consuming. Natural Language Processing (NLP) is the sub-division of Artificial Intelligence that narrows down the gap between technology and human cognition by extracting the relevant information from the pile of data. In the present work, scientific information regarding the Friction Stir Welding of Aluminum alloys was collected from the abstract of scholarly research papers. For extracting the relevant information from these research abstracts four Natural Language Processing based algorithms i.e. Latent Semantic Analysis (LSA), Luhn Algorithm, Lex Rank Algorithm, and KL-Algorithm were used. In order to evaluate the accuracy score of these algorithms, Recall-Oriented Understudy for Gisting Evaluation (ROUGE) was used. The results showed that the Luhn Algorithm resulted in the highest f1-Score of 0.413 in comparison to other algorithms.

## 1. Introduction

Artificial intelligence (AI) is a broad field of computer science that focuses on creating intelligent machines that can accomplish activities that would normally need human intelligence. Artificial intelligence can take data from sensors, equipment, and people, then apply it to algorithms to improve operations or accomplish light-out production. AI-driven robots are paving the way to the future by providing a slew of benefits, including new prospects, increased production efficiency, and a tighter match between machine and human interaction [1-6]. The Fourth Industrial Revolution is characterized by knowledge-based work that is carried out through automation and the development of new techniques to automate jobs.

There has been a significant amount of study into the intellectualization of industrial control, which is currently rather mature. However, industrial data intellectualization is still in its early stages, and there are some challenges, primarily because industrial data is heterogeneous and multisource, and the majority of it is unstructured data. As a result, figuring out how to automatically extract usable information from these unstructured data and integrate it is an important aspect of industrial data intelligence. Services centered on the users' experience, for example, are a significant aspect of value generation and are becoming increasingly relevant in the vehicle sector.

Natural Language Processing (NLP) is a computational technique that is a branch of Artificial Intelligence that gives the ability to a machine system to understand the audible words and text just like humans [7-8]. Natural Language Processing algorithms work by separating the human-sourced language into fragments for understanding the given context by analyzing the main ideas conveyed by the words and further analyzing the grammatical structure of the input sentences [9-10]. Natural Language Processing finds application in monitoring of social media, language translation, chatbots, grammatical corrections, targeted population advertisement, etc [11].

Natural Language Processing is also been applied to the material science and manufacturing domain for knowledge extraction purposes. Shetty et al. [12] used Natural Language Processing algorithm for tracking the popularity of polymers on the basis of their applications and also for the determination of the polymer material for novel applications as shown in Figure 1. Venugopal et al. [13] developed a framework for automating image comprehension and text in order to extract the knowledge from the available works of literature on inorganic glass materials.

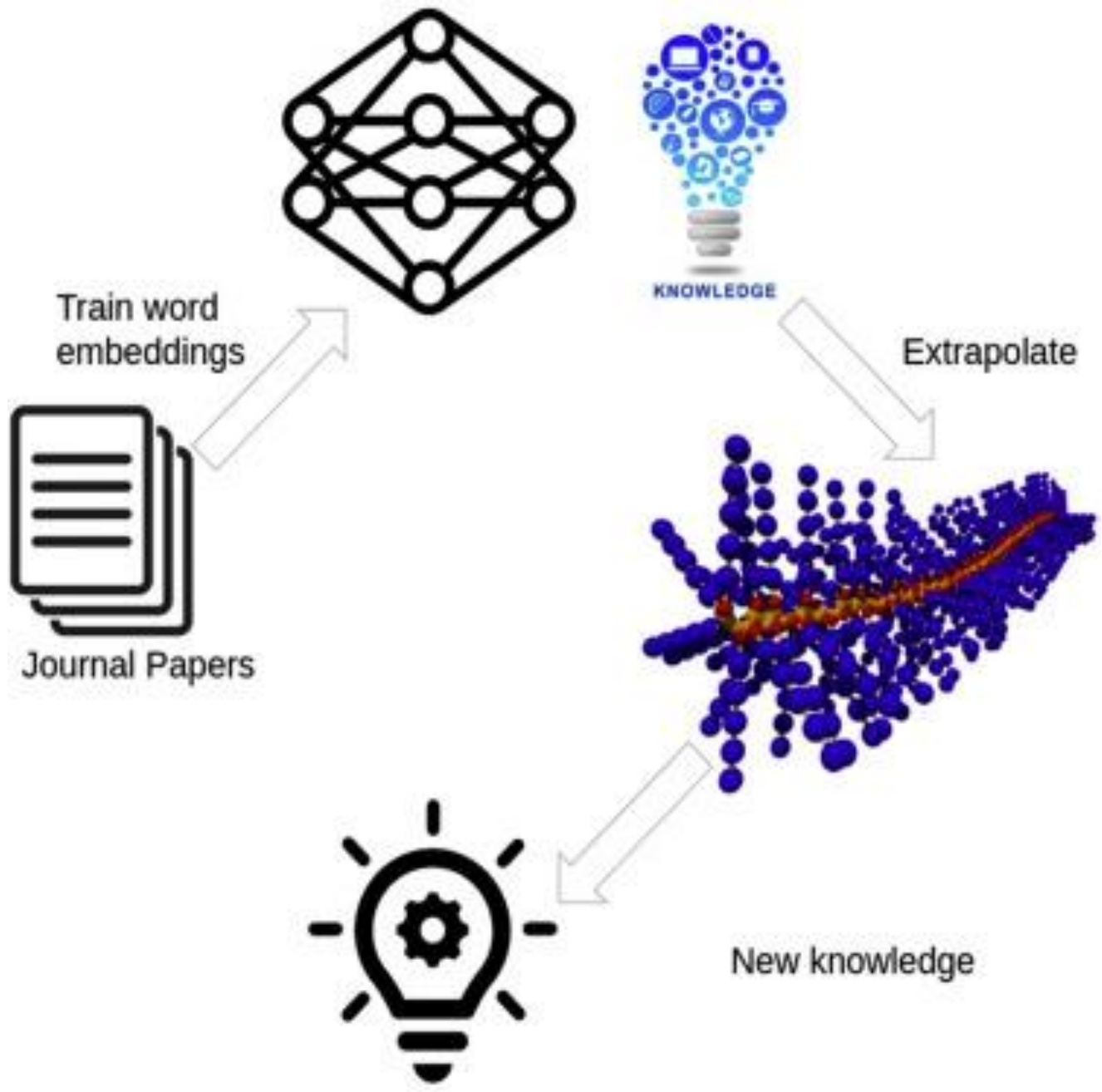

Figure 1: Implementation of Natural Language Processing algorithm on Polymers' literatures [6].

There are limited papers available on the application of Natural Language processing in the material science and manufacturing domain. The present research work is the first time application of Natural Language Processing in the Friction Stir Welding domain for knowledge extraction purposes. The main objective of the research work is to summarize the research study carried out in the field of Friction Stir Welded aluminum joints by using various algorithms and further finding out the best algorithm for summarization purposes.

## 2. Experimental Procedure

Firstly, 20 abstracts dealing with the Friction Stir Welded aluminum joints were collected from the Google Scholar database [14–32]. The collected abstracts were converted into a text file for further processing as shown in Figure 2. Secondly, the sentences in the given text file were tokenized and frequency distribution was calculated. Thirdly, the punctuation marks in the given text were removed followed by the finding and removal of the stop words in the text file. Fourthly, the obtained text was subjected to 4 Natural Language Processing based algorithms, and their metric features such as recall, precision, and f1-score values were calculated by using the ROUGE-N algorithm. N represents the n-gram which is a group of tokens or words.

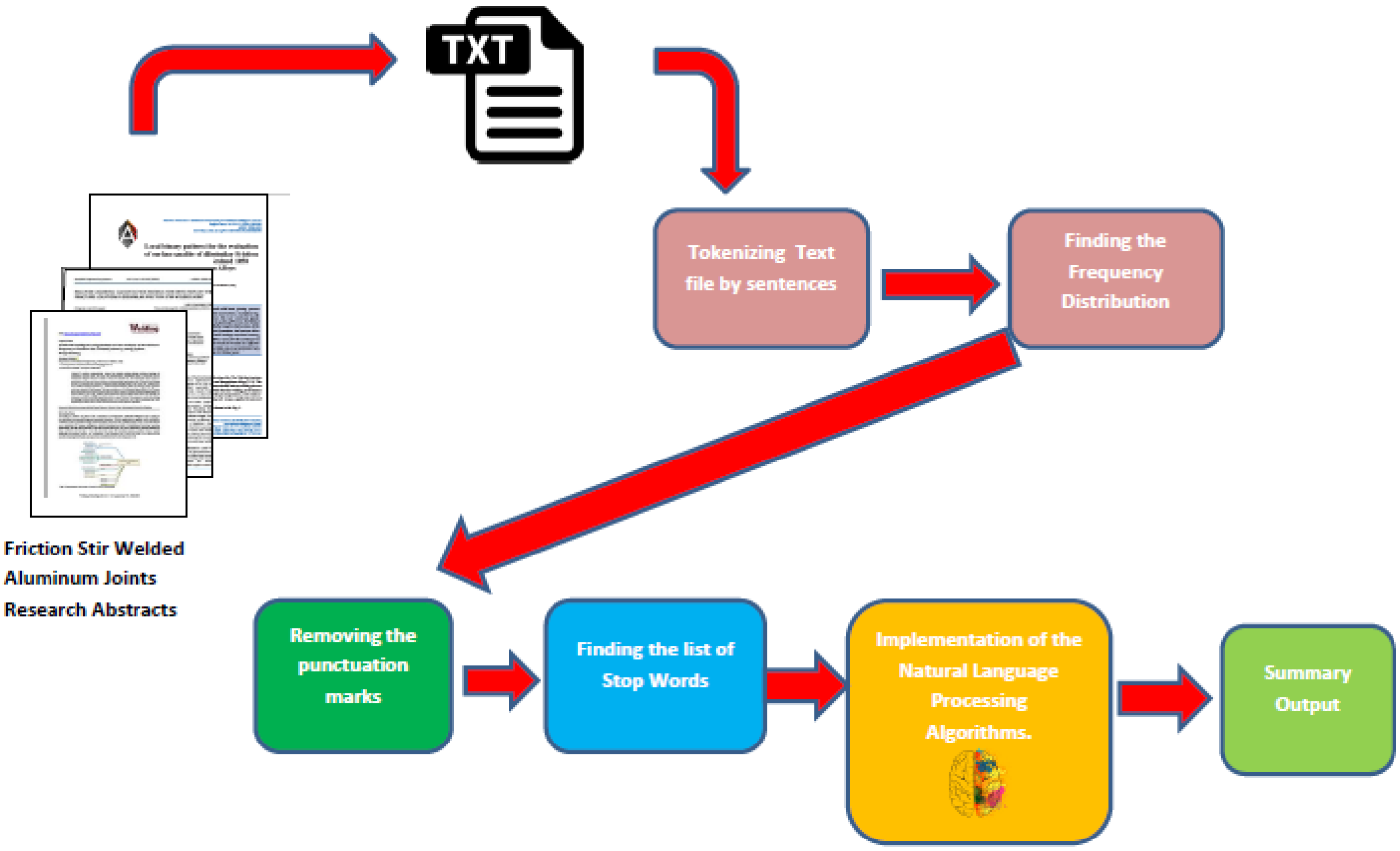

Figure 2: Implementation of Natural Language Processing Algorithm in recent work.

The calculation of the recall value, precision value, and f1-score is done by Equation 1, Equation 2, and Equation 3.

$$Recall\ value = \frac{number\ of\ n{-}grams\ found\ in\ model\ and\ reference}{number\ of\ n{-}grams\ in\ reference} \quad (1)$$

$$Precision\ value = \frac{number\ of\ n{-}grams\ find\ in\ model\ and\ reference}{number\ of\ n{-}grams\ in\ model} \quad (2)$$

$$F1 - Score = 2 \times \frac{Precsion\ Value \times Recall\ Value}{Precision\ Value + Recall\ Value} \quad (3)$$

**3. Results and Discussion**

The initial plot of the frequency distribution is shown in Figure 3. The final plot of the frequency distribution after removing the punctuation marks and stop words is shown in Figure 4. The word cloud obtained after these pre-processing steps is shown in Figure 5.

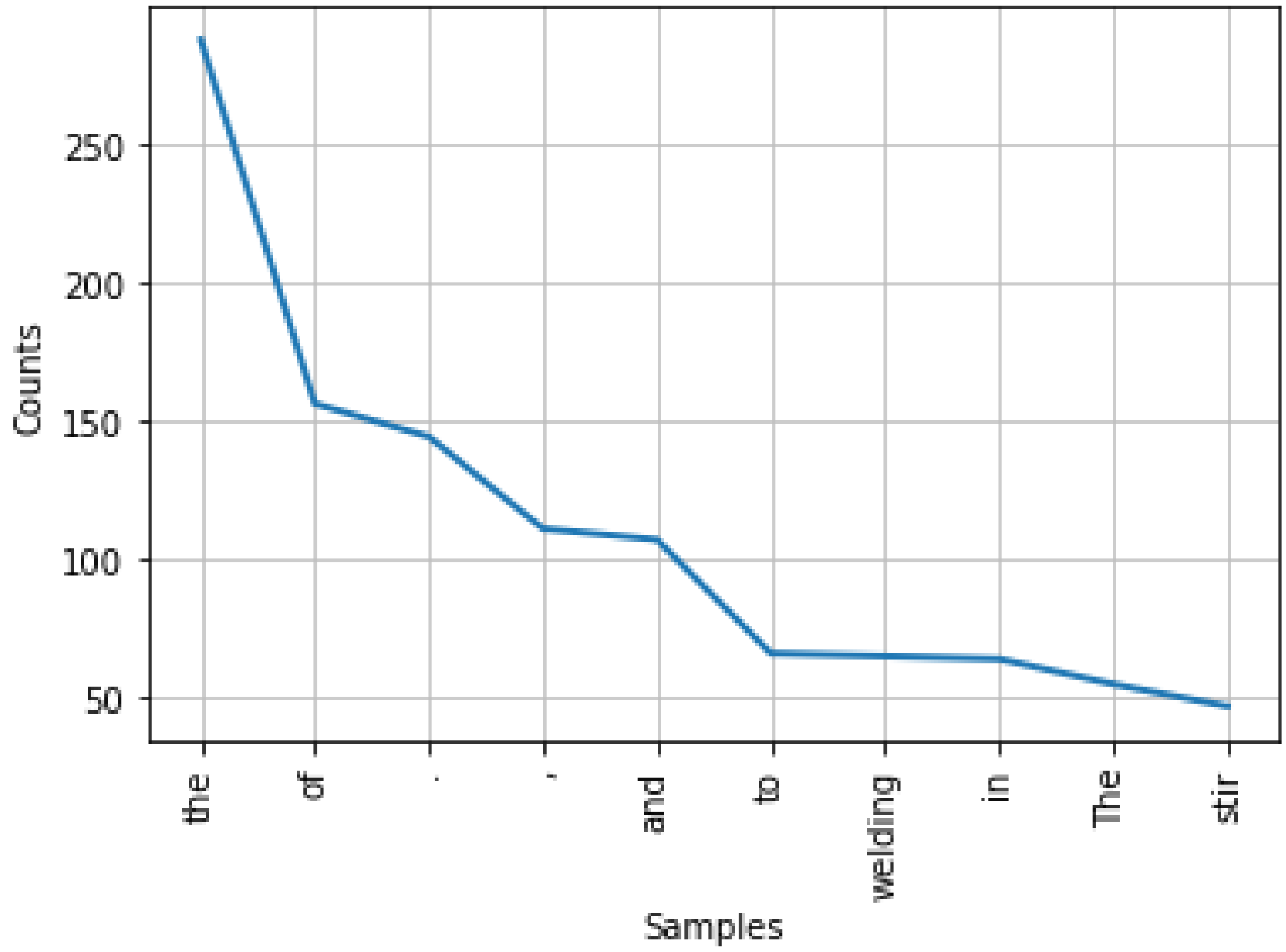

Figure 3: Frequency Distribution plot at an initial stage

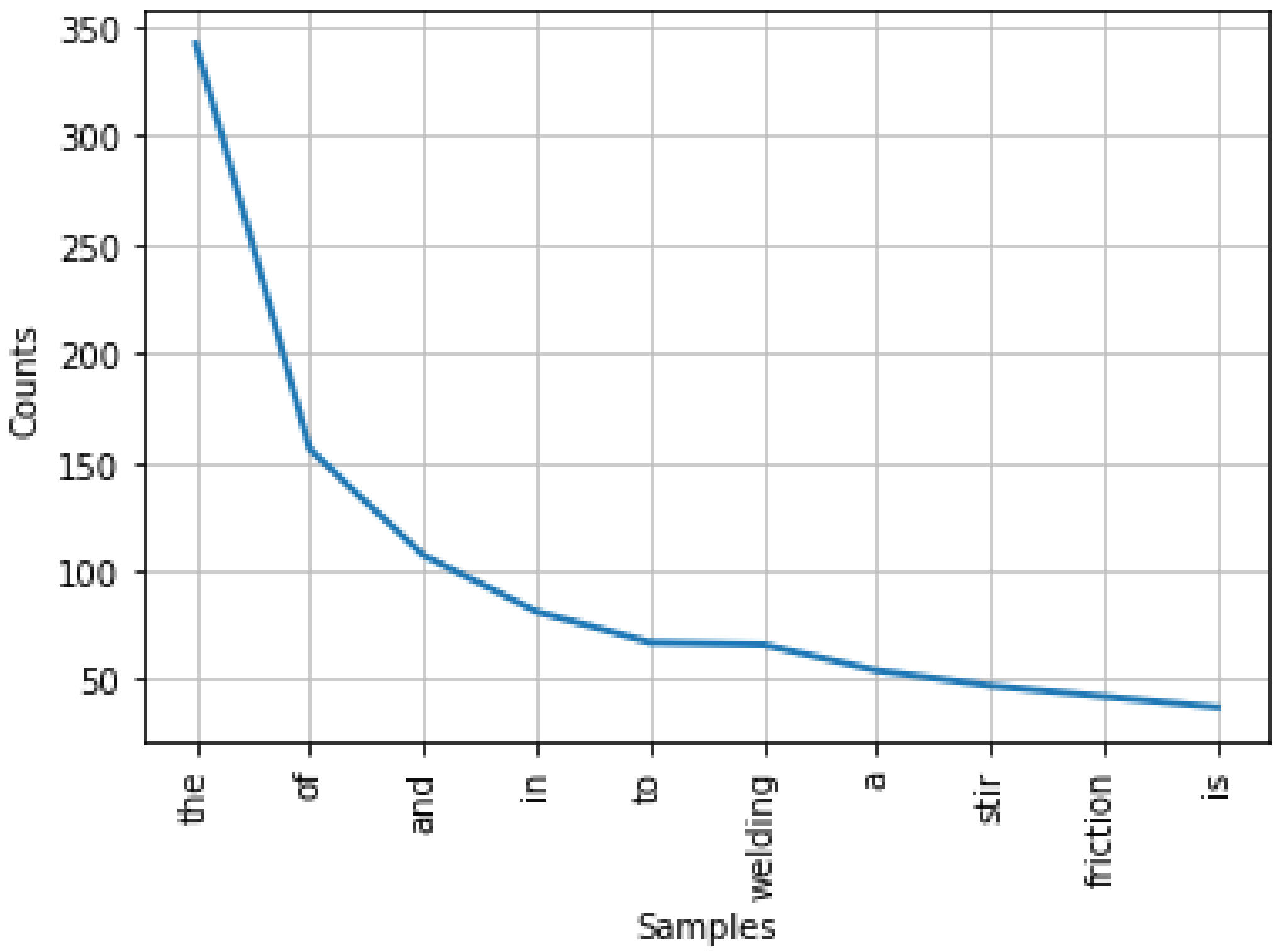

Figure 4: Final plot of frequency distribution after removing the punctuation marks and stop words.

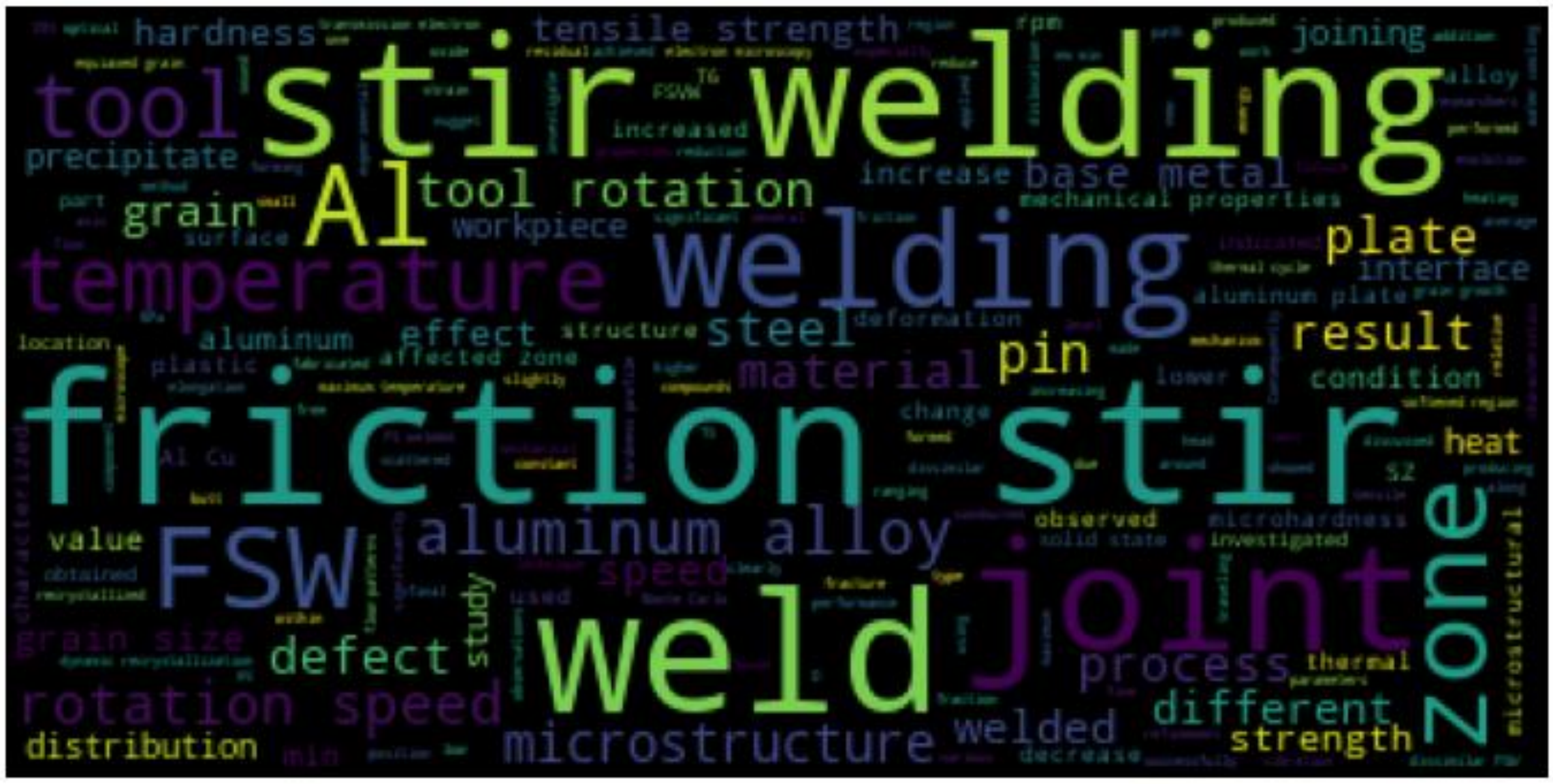

Figure 5: Word Cloud obtained for the given text file

### 3.1 Text Summarization by LexRank Algorithm

LexRank algorithm is an unsupervised eigenvector centrality-based graphical method to summarize the given input text automatically. For a graphical representation of a required sentence, a connectivity matrix that works on the cosine similarity principle is implemented as an adjacency matrix. In this approach, a centroid sentence is chosen which acts as a mean for all other sentences present in the text file which is followed by the ranking allocation on the basis of their similarities. The bag of words model is used to define similarity by representing the N (number of all possible words)- dimensional vectors. In the vector representation form of the sentence, the value of the corresponding dimension is calculated by Equation 4.

$$Value\ of\ corresponding\ dimension = Inverse\ Document\ Frequency\ (IDF)\ of\ the\ word \times number\ of\ occurrences\ of\ the\ word \quad (4)$$

The IDF modified cosine value of the word is calculated by Equation 5.

$$idf - modified - cosine(x,y) = \frac{\sum_{w \in x,y} tf_{w,x} tf_{w,y} (idf_w)^2}{\sqrt{\sum_{x_i \in x} (tf_{xi,x} idf_{xi})^2} \times \sqrt{\sum_{yi \in y} (tf_{yi,y} idf_{yi})^2}} \quad (5)$$

In order to display the output summary, number of sentence were equated to 11. The output obtained is shown below:

*The authors tried to butt-weld an aluminum alloy plate to a mild steel plate by friction stir welding, and investigated the effects of a pin rotation speed, the position for the pin axis to be inserted on the tensile strength and the microstructure of the joint. The maximum tensile strength of the joint was about 86% of that of the aluminum alloy base metal. In the study, the microstructural evolution and mechanical properties of friction stir welding joint during plastic forming and subsequent heat treatment were investigated. The gain size in the SZ is smaller than that in the base metal, and is decreased with a decrease of the tool rotation speed. Friction stir welding (FSW) of 2219-O and 2219-T6 aluminum alloys was performed to investigate the effects of the base material conditions on the FSW characteristics. The experimental results indicated that the base material condition has a significant effect on weld morphologies, weld defects, and mechanical properties of joints. However, the grain size of the stir zone decreased with the decrease in friction heat flow during FSW. The aluminum (Al) alloys 6063-T5 and T4 were friction-stir welded at different tool rotation speeds (R), and then distributions of the microstructure and hardness were examined in these welds. This study aims to experimentally explore the thermal histories and temperature distributions in a workpiece during a friction stir welding (FSW) process involving the butt joining of aluminum 6061-T6. The dynamic recrystallization (DRX) process and grain structure evolution during friction stir welding determine the final microstructure and*

*performance of weld joints. The joint efficiency (the ratio of the joint strength to the base metal strength) of the joint fabricated by FSVW with water cooling was around 87%, while this quantity was about 66% for the friction stir welding joint.*

Table 1 shows the recall value, precision value and f1-score obtained by using the LexRank algorithm.

Table `1: Performance evaluation of LexRank algorithm

| Recall Value | Precision Value | F1-Score |
|---|---|---|
| 1.0 | 0.144 | 0.252 |

So, it is observed that the LexRank is an unsupervised text summarizing method based on graph-based sentence centrality rating. The essential premise is that sentences "recommend" to the reader more similar ones. As a result, if one sentence is highly similar to many others, it is likely to be a very important sentence. This sentence's relevance arises from the importance of the sentence "recommending" it. As a result, in order to be highly ranked and included in a summary, a sentence must be similar to a large number of other sentences. This makes obvious sense and enables the algorithms to be used on any fresh text.

### 3.2 Text Summarization by Latent Semantic Analysis (LSA) Algorithm

Latent Semantic Analysis (LSA) algorithm also known as Latent Semantic Indexing (LSI) is used for simulating and modeling the essence of passages and words given in the text document by using Singular Value Decomposition (SVD) method. Latent Semantic Analysis algorithm statistically analyze and predict the meaning of a given passage by analyzing the co-occurance of the given word in a text file. The summary obtained by implementing the Latent Semantic Algorithm (LSA) on a given experimental text file is shown below:

*In evaluating friction stir welding, critical issues (beyond a sound joint) include microstructure control and localized mechanical property variations. Being a solid-state process, friction stir welding has the potential to avoid significant changes in microstructure and mechanical properties. Dissimilar friction stir welding between magnesium and aluminum alloy plates with thicknesses of 2 mm was performed. The tool for welding was rotated at speeds ranging from 800 to 1600 rpm under a constant traverse speed of 300 mm/min. Friction stir welding is an efficient manufacturing method for joining aluminum alloy and can dramatically reduce grain size conferring excellent plastic deformation properties. The tool for welding was rotated at speeds ranging from 500 to 3?000 r/min under a constant traverse speed of 100 mm/min. In this study, an attempt has been made to investigate the formation of these defects. Precipitate evolution in friction stir welding of 2219-T6 aluminum alloys was characterized by transmission electron microscopy. FSW is a new solid-state welding process used to join metals and alloys (especially unweldable*

*aluminums). The friction-stir welding (FSW) of 0.6 cm plates of 2024 Al (140 HV) to 6061 Al (100 HV) is characterized by residual, equiaxed grains within the weld zone having average sizes ranging from 1 to 15 ?m, exhibiting grain growth from dynamically recrystallized grains which provide a mechanism for superplastic flow; producing intercalated, lamellar-like flow patterns. The hardness values increased from 50 to around 78 (Hv) for friction stir welding and FSVW with water systems, respectively.*

The performance evaluation results of LSA algorithm is shown in Table 2:

Table `2: Performance evaluation of LSA Algorithm

| **Recall Value** | **Precision Value** | **F1-Score** |
|---|---|---|
| 0.987 | 0.155 | 0.268 |

It is observed that the recall value and precision value of LSA algorithm is less than those of LexRank algorithm but f1-score of LSA algorithm is more than those of LexRank algorithm.

It is can be seen that the Latent Semantic Analysis (LSA) is a mathematical technique that is used to get insight into a material. Topic Modeling is based on this method. The basic concept is to decompose a matrix of what we have i.e. documents and terms into two independent document-topic and topic-term matrices.

### 3.3 Text Summarization by using Luhn Algorithm

Luhn algorithm is based on heuristic approach and is one of the earliest method used for summarizing the text file. The mechanism of the Luhn algorithm is based on the selection of higher importance words on the basis of their frequency distribution and further assigning higher weights to the words which are mentioned in the beginning of the text file. The score assigned to a sentence is calculated by Equation 6.

$$Score = \frac{(Number\ of\ meaningful\ words)^2}{Span\ of\ meaningful\ words} \qquad (6)$$

The output summary obtained by the given algorithm is as follows:

*Simulated weld thermal cycles with different peak temperatures have shown that the precipitates are dissolved at temperatures higher than 675 K and that the density of the strengthening precipitate was reduced by thermal cycles lower than 675 K. A comparison between the thermal cycles and isothermal aging has suggested precipitation sequences in the softened region during friction-stir welding. The authors tried to butt-weld an aluminum alloy plate to a mild steel plate by friction stir welding, and investigated the effects of a pin rotation speed, the position for the pin axis to be inserted on the tensile strength and the*

*microstructure of the joint. Many fragments of the steel were scattered in the aluminum alloy matrix and the oxide film removed from the faying surface of the steel by the rubbing motion of a rotating pin was observed at the interface between the steel fragments and the aluminum alloy matrix. Tensile strength of the welds is about 20% lower than that of the hardening aluminum plate, but about 10% higher microhardness is demonstrated by the welds in comparison with that of the aluminum plate in annealing condition. Moreover, travel rate of the welding head pin has a strong effect on microhardness and tensile strength of the FSW welds, and the ratio of rotation speed and travel rate of the head should be in a reasonable range to obtain high performance welds. Different R values did not result in significant differences in the hardness profile in these welds, except for the width of the softened region in the weld of 6063-T5 Al. Postweld aging raised the hardness in most parts of the welds, but the increase in hardness was small in the stir zone produced at the lower R values. The welding of a lap joint of a commercially pure aluminum plate to a low carbon steel plate (i.e., Al plate top, and steel plate bottom) was produced by friction stir welding using various rotations and traveling speeds of the tool to investigate the effects of the welding parameters on the joint strength. The friction-stir welding (FSW) of 0.6 cm plates of 2024 Al (140 HV) to 6061 Al (100 HV) is characterized by residual, equiaxed grains within the weld zone having average sizes ranging from 1 to 15 ?m, exhibiting grain growth from dynamically recrystallized grains which provide a mechanism for superplastic flow; producing intercalated, lamellar-like flow patterns. Dislocation spirals and loops are also observed in the 2024 Al intercalation regions within the weld zones at higher speeds (>800 rpm) corresponding to slightly elevated temperatures introducing dislocation climb, and residual microhardness profiles follow microstructural variations which result in a 40% reduction in the 6061 Al workpiece microhardness and a 50% reduction in the 2024 Al workpiece microhardness just outside the FSW zone. It is found that along the material flowing path during friction stir welding, the main DRX mechanisms are different at leading, retreating and trailing sides around the tool, and the tool rotation speed (associated with the heat input level) determines which type of DRX is easier to occur at different positions along material flowing path. The joint efficiency (the ratio of the joint strength to the base metal strength) of the joint fabricated by FSVW with water cooling was around 87%, while this quantity was about 66% for the friction stir welding joint.*

The result of performance evaluation metrics is shown in Table 3.

Table 3: Performance evaluation of Luhn Algorithm

| Recall Value | Precision Value | F1-Score |
|---|---|---|
| 0.996 | 0.260 | 0.413 |

It is observed that the f1-score of the Luhn Algorithm is greater than the LexRank and LSA algorithm. And also it can be seen that the Luhn's algorithm is a TF-IDF-based technique. Only the most important terms are selected based on their frequency. The words that appear at the start of the document are given more weight.

### 3.4 Text Summarization by using KL-Algorithm

Kullback-Lieber (KL) algorithm is based on greedy method approach by adding a sentence to the summary when there is a reduction in KL divergence. KL algorithm checks the divergence of the summary vocabulary from its input vocabulary by minimizing it. If there are L words in a summary and for a given document D, a criterian is introduced by KL-algorithm for selection of summary S. The output result obtained by the given algorithm is as follows:

*The maximum tensile strength of the joint was about 86% of that of the aluminum alloy base metal. The elongation of the FSWed plates is lower than that of the base metal (about 22%). However, it is noticeable that the maximum elongation of about 21% is obtained at 1?000 r/min. In addition, the two types of joints have different fracture location characteristics. However, the origin of the defects remains an area of uncertainty. In the heat-affected zone the precipitates coarsened. Tensile strength of the welds is about 20% lower than that of the hardening aluminum plate, but about 10% higher microhardness is demonstrated by the welds in comparison with that of the aluminum plate in annealing condition. However, the grain size of the stir zone decreased with the decrease in friction heat flow during FSW. The ductility in FS-welded 5083 Al alloy increased with the decrease in friction heat flow. It was indicated that the formability in FS-welded 5083 Al alloy was improved by the refinement of grain size of the stir zone. Microstructural analyses suggested that the small increase in hardness in the stir zone produced at the lower R values was caused by an increase in the volume fraction of PFZs.*

Table 4 shows the result of the performance of KL- algorithm. It is observed that the f1-score of KL-algorithm which is lowest of all the implemented algorithms.

Table 4: Performance evaluation of KL-Algorithm

| **Recall Value** | **Precision Value** | **F1-Score** |
|---|---|---|
| 0.990 | 0.100 | 0.182 |

## 4. Conclusion

In order to create an autonomous text summary, machine learning and natural processing language come in handy. In machine-human interaction, natural language processing plays a significant role. It is still being worked on, and more study is being done in this area. In the future, we might witness a more intelligent and perfect text summary system that understands human language and works appropriately. In the present research work, four types of Natural

Language Processing algorithms i.e. LexRank algorithm, Luhn algorithm, Latent Semantic Analysis (LSA) algorithm were successfully implemented on the research texts summarization of Friction Stir Welded Aluminum joints. The results showed that the Luhn algorithm yielded the highest f1-score of 0.413 in comparison to the other algorithms. It can be also concluded that the text mining approach can be beneficial for the researchers to mine out the necessary information from the bulk of data by automating the whole process which leads to the reduction of the experimentation time.